\documentclass[runningheads]{llncs}
\usepackage{graphicx}
\usepackage{enumerate}
\usepackage[table]{xcolor}
\usepackage{bm}
\usepackage{multirow}
\usepackage{todonotes}
\usepackage{soul}
\usepackage{subfig}
\usepackage{wrapfig}

\makeatletter
\RequirePackage[bookmarks,unicode,colorlinks=true]{hyperref}%
   \def\@citecolor{blue}%
   \def\@urlcolor{blue}%
   \def\@linkcolor{blue}%

\def\orcidID#1{\smash{\href{http://orcid.org/#1}{\protect\raisebox{-1.25pt}{\protect\includegraphics{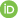}}}}}
\makeatother

\begin{document}
\title{Assessing the Understandability and Acceptance of Attack-Defense Trees for Modelling Security Requirements}
\titlerunning{Assessing the Understandability and Acceptance of Attack-Defense Trees}

\author{Giovanna Broccia\inst{1}\orcidID{0000-0002-4737-5761} 
\and
Maurice H. ter Beek\inst{1}\orcidID{0000-0002-2930-6367} 
\and \\
Alberto Lluch Lafuente\inst{2}\orcidID{0000-0001-7405-0818}
\and
Paola Spoletini\inst{3}\orcidID{0000-0001-7922-4936}
\and
Alessio Ferrari\inst{1}\orcidID{0000-0002-0636-5663} 
}
\authorrunning{G. Broccia et al.}
\institute{ISTI-CNR, Pisa, Italy\\
\email{\{giovanna.broccia, maurice.terbeek, alessio.ferrari\}@isti.cnr.it}
\and
DTU, Lyngby, Denmark\\
\email{albl@dtu.dk} 
\and
Kennesaw State University, GA, USA\\
\email{pspoleti@kennesaw.edu}
}
\maketitle              % typeset the header of the contribution
\begin{abstract}
\textit{Context and Motivation} Attack-Defense Trees (ADTs) are a gra\-phi\-cal notation used to model and assess security requirements. ADTs are widely popular, as they can facilitate communication between different stakeholders involved in system security evaluation, and they are formal enough to be verified, e.g., with model checkers.
\textit{Question/Problem} While the quality of this notation has been primarily assessed quantitatively, 
its understandability has never been evaluated despite being mentioned as a key factor for its success.
\textit{Principal idea/Results} In this paper, we conduct an experiment with 25 human subjects to assess the understandability and user acceptance of the ADT notation. The study focuses on performance-based variables and perception-based variables, with the aim of evaluating the relationship between these measures and how they might impact the practical use of the notation. The results confirm a good level of understandability of ADTs. Participants consider them useful, and they show intention to use them.  \textit{Contribution} This is the first study empirically supporting the understandability of ADTs, thereby contributing to the theory of security requirements engineering.

\end{abstract}

\section{Introduction}
The definition of security requirements entails the representation and analysis of envisioned threats and mitigation solutions, oriented to eventually define a security policy~\cite{fabian2010comparison}. Several notations have been proposed in requirements engineering (RE) to model and analyse security requirements, such as extensions of well-known notations (e.g., Secure~I$^*$~\cite{liu2009secure} and Secure UML~\cite{lodderstedt2002secureuml}) and other comprehensive notations with analysis capabilities (e.g., the Socio-Technical Security Modelling Language (STS-ML)~\cite{paja2015modelling} and the Restricted Misuse Case Modeling (RMCM) approach~\cite{MAI2018165}).

Among this variety of proposals, Attack-Defense Trees (ADTs) offer a graphical notation used to model and assess the security requirements of systems or assets. They provide a representation of possible actions an attacker might take to attack a system and the measures that a defender can employ to protect the system~\cite{kordy2011foundations}. 
The purposes of ADTs are multiple. In addition to providing a threat modelling methodology, they can be used for quantitatively assessing the security of a system (e.g., with model checking). Moreover, ADTs are useful 
for facilitating communication between stakeholders from different fields and with different backgrounds (e.g., domain experts, security experts). 

Several studies have shown how graphical notations 
are more comprehensible by humans than textual notations~\cite{sharafi2013empirical,stein2005graphical}. 
However, although ADTs have been claimed as one of the most popular graphical models for system security analysis~\cite{gadyatskaya2018new}, extremely easy to use also for novice users~\cite{widel2019beyond}, and as an easily understandable human-readable notation~\cite{eisentraut2021assessing}, no user study has been proposed to verify these hypotheses. Albeit this research direction holds promise and would be helpful in evaluating their effectiveness \cite{gadyatskaya2018new,8101532,lallie2020review}. Indeed, beyond the realm of attack trees, there exists a substantial body of empirical research literature focused on security modelling and assessment \cite{4159906,labunets2017equivalence,6681349}.
These kinds of studies are particularly beneficial given the centrality of humans in system security---both for possible insider attacks and for human errors that make the system vulnerable \cite{eisentraut2021assessing}. 

In this paper, we present the first experiment that aims at 
investigating the quality of the ADT notation, both in terms of understandability and in terms of 
user acceptance. 
We designed the study based on the Method Evaluation Model (MEM)~\cite{moody2001dealing}, a model used to evaluate information technologies, which extends the Technology Acceptance Model (TAM)~\cite{davis1989perceived}.
We adapt MEM following the approach by Abrah{\~a}o~\cite{abrahao2011evaluating} and identify two classes of variables: performance-based and perception-based.
The performance-based variables aim at assessing the understandability of ADTs, while perception-based variables seek to evaluate the users' acceptance of ADTs.

Our results
show that: (1)~ADTs are sufficiently understandable; (2)~ADTs are perceived as easy to use and useful, and participants express the intention to use them; (3)~there is a relationship between perceived usefulness and intention to use; (4)~there are no significant relationships between various performance-based measures of understandability (effectiveness and efficiency) and perception-based variables (ease of use, usefulness, intention to use), except in the following cases: (a) perceived ease of use has a positive relationship with effectiveness, i.e., those who make fewer mistakes in different ADT understandability tasks generally consider the notation easier; (b) those who \textit{apply} the method better in practice also consider it more useful; (c) those who make fewer mistakes when \textit{observing} the notation used in realistic contexts, consider the method easier. 
Our replication package is publicly available~\cite{zenodo10067064}. 

%\smallskip
%\noindent
\paragraph{Related Work.}  Several notations have been proposed in RE to model and analyse security requirements~\cite{mellado2010systematic,iankoulova2012cloud,souag2016reusable,villamizar2018systematic}. Some of these notations are extensions of existing notations, like Secure~I$^*$~\cite{liu2009secure}, KAOS~\cite{salehie2012requirements}), Secure UML~\cite{lodderstedt2002secureuml}, Misuse cases~\cite{sindre2005eliciting}, and Secure Tropos~\cite{giorgini2007modelling}.

Other attempts, some based on the languages above, also offer analysis capability.
In particular, the ones mentioned in the Introduction. STS-ML~\cite{paja2015modelling} is an actor- and goal-oriented security requirements modelling language based on Tropos, able to capture system security needs and requirements at the organisational level and reason about corporate assets, social dependencies, and trust properties. 
RMCM~\cite{MAI2018165} is a use case-driven modelling method that uses misuse case diagrams~\cite{sindre2005eliciting} to support the specification of security and privacy requirements of multi-device software ecosystems in a structured and analysable form.
 The Risk-based Security Requirements (RBSR) model~\cite{Ezenwoye22risk} associates security requirements with specific weaknesses and risk profiles that can vary over time and provides mitigation accordingly to these variations. 
Finally,~\cite{zareen2020security} introduces a threat-based security framework and its Business Process Model and Notation (BPMN) extension to model the security threat and support risk analysis.

Labunets et al. observed a difference in the representation of security risk assessment between academic proposals and industry standards. Academic approaches favour graphical notation, while the industry leans towards tabular models. Several studies were conducted to compare the effectiveness of graphical and tabular models. \cite{labunets2017equivalence} proved that both methods are equally effective. In ~\cite{6681349}, a comparative analysis of visual and textual risk-based approaches revealed that the visual method is more effective for identifying threats, while the textual method is slightly better for eliciting security requirements.

In~\cite{8101532}, the results of an empirical evaluation conducted to determine the effectiveness of two attack modelling techniques, an adapted attack graph method and the fault tree standard, are reported. 
%The objective was to test the participants' ability to recall, comprehend, and apply these techniques. 
The results indicate that the attack graph method is more effective than the fault tree method. %Furthermore, participants with a computer science background performed better than those without when using both methods. The study emphasises the need for further comparisons in a wider range of settings involving additional techniques.

%In~\cite{4159906}, a thorough assessment and comparison of a selected group of risk analysis techniques based on a practical case study was presented.
%The objective of the study was to measure a set of high-level criteria, comparing the raw output of different methods, and using statistical techniques to assess the overall performance of each method. The analysis led to several observations regarding the quality and categorisation capabilities of the techniques.

%In~\cite{hogganvik2006graphical}, a graphical approach to facilitate communication and understanding among different classes of users during a risk analysis brainstorming session was proposed. The development and the guidelines for the use of such a graphical language were based on a combination of empirical investigations and experiences gathered from utilising the approach in large-scale industrial field trials by both professionals and students.

\section{Attack-Defense Trees}
\label{sec:ADTs}

The assessment of system security through graphical tree structures originated in 1960 with fault tree analysis~\cite{vesely1981fault}, and gradually spread with the usage of similar structures such as attack trees~\cite{schneier1999attack,dh2005foundations}.
To manage the dynamic nature of system security, 
Attack-Defense Trees (ADTs)~\cite{kordy2011foundations} were introduced, extending attack trees with defense strategies and quantitative risk assessment~\cite{kordy2013ADTool,terbeek2021quantitative}. 
ADTs model attack-defense scenarios, namely $2$-player games between a proponent and an opponent.

Formally, ADTs are rooted trees with labelled nodes
of two opposite types: attack nodes and defense nodes, representing the goals of the attacker and the defender, respectively. 
The root can be either type: if the root is an attack node, the proponent is an attacker; conversely, if the root is a defense node, the proponent is a defender.
The main goal can be refined into sub-goals, described by its child nodes of the same type. The refinement can be either conjunctive 
(i.e., all sub-goals must be achieved to achieve the parent goal) or disjunctive 
(i.e., at least one of the sub-goals must be achieved to reach the parent goal). A node with no children of the same type is called a non-refined node, and it represents a basic/atomic action.
Each node may have one child of the opposite type, representing a countermeasure to its (sub-)goal. Essentially, an attack node may have a number of children that refines the attack and a single defense node that fends it off. Conversely, a defense node may have a number of children which refines the defense, and a single attack node that counterattacks it. 
\begin{wrapfigure}{r}{0.45\textwidth}
\vspace{-0.5cm}
\centering
\includegraphics[width=0.40\textwidth]{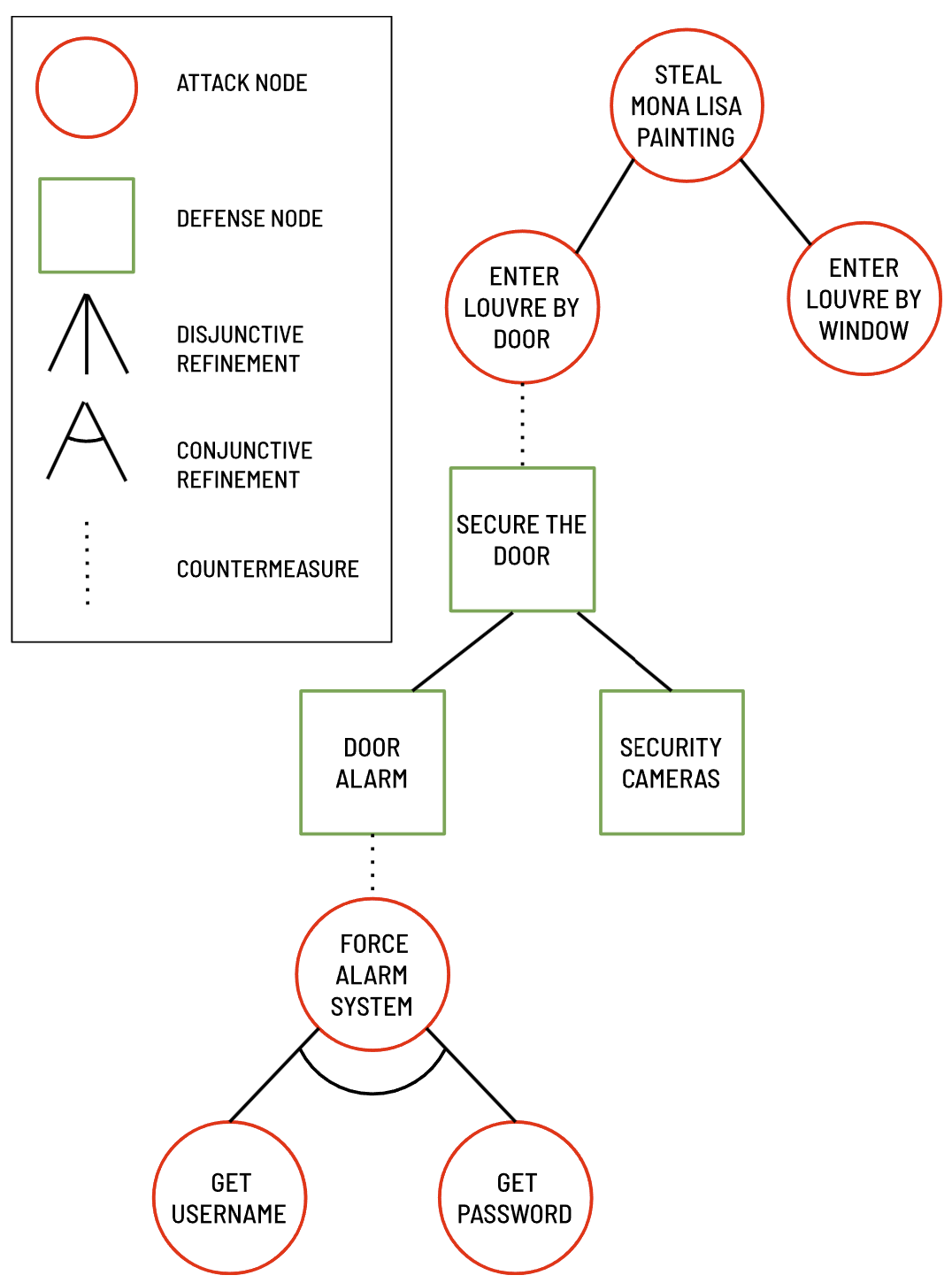}
\caption{ADT for theft of Mona Lisa.} \label{fig:ADT}
\vspace{-0.5cm}
\end{wrapfigure}

To demonstrate the features of ADTs, we present a simple fictitious scenario describing the theft of the Mona Lisa painting (cf.\ Fig.~\ref{fig:ADT}). To steal the painting, two kinds of attacks can be carried out: enter the Louvre museum by the door or by the window. Figure~\ref{fig:ADT} shows in detail only the door branch (further attacks and defenses could easily be added). To secure the door, the museum can use an alarm; however, the attacker can perform a counterattack by forcing the alarm system. To do so, the attacker needs to get both the username and the password.

Evaluation of ADTs has so far considered issues like the consistency between an ADT and the system and the impact of repeated labels on results~\cite{audinot2017my,kordy2018quantitative}.
As far as we know, there is no work in the literature that has focused on the assessment of the comprehensibility of ADTs (neither of attack trees). Albeit their comprehensibility is usually assessed as a factor of success~\cite{eisentraut2021assessing,widel2019beyond,gadyatskaya2018new}.

\section{Method Evaluation and Technology Acceptance Model}
\label{sec:model}

The Method Evaluation Model (MEM)~\cite{moody2001dealing} is a model used to evaluate new information technologies.  
According to MEM, the usage of new technologies is influenced by a set of \textit{perception-based} variables and \textit{performance-based} variables.

The perception-based variables are used to gauge the level of \textit{acceptance} of the technology and include the perceived ease of use (PEOU), which measures how easy the technology is perceived to be, the perceived usefulness (PU), which measures how useful the technology is perceived to be, and the intention to use (ITU), which measures the extent to which users intend to use the technology in the future.
The performance-based variables consist of efficiency and effectiveness, which measure the effort required to use the technology and how well the technology has been used to reach the goals, respectively.
Essentially, the adoption of a new technology depends not only on whether it is actually effective but also on whether the users perceive it to be effective.

MEM has been applied in the fields of RE~\cite{abrahao2011evaluating} and language comprehension~\cite{broccia23neverlang}.
In both studies, the performance-based variables (efficiency and effectiveness) 
have been adapted to measure the \textit{understandability} of requirement models and language constructs, respectively. In practice, the performance-based variables are understandability effectiveness and understandability efficiency, computed based on the results obtained by sample subjects in problem-solving tasks. This paper adopts this approach and further decomposes the variables into fine-grained dimensions (cf. Sect.~\ref{sec:stDes}). In line with MEM, we evaluate if these variables are related to perception-based variables. 
\enlargethispage*{1.1\baselineskip}

\section{Study Design}\label{sec:stDes}

Our experiment aims to study the degree of ADTs understandability
and users' acceptance. We also study if there is a relationship between the degree of acceptance of the notation and its understandability.

\subsection{Variables, Research Questions, and Tests}

\paragraph{Acceptance and Understability Dimensions.}
Users' acceptance is based on the MEM model presented in Section~\ref{sec:model}. In particular, we evaluate \textit{acceptance} using the three perception-based variables from the MEM (PEOU, PU, and ITU). 

\textit{Understandability} is evaluated in terms of effectiveness and efficiency based on the results of sample subjects in some problem-solving tasks (as suggested by the literature, e.g., \cite{oliveira2020evaluating}). For both effectiveness and efficiency, we further distinguish between fine-grained understandability, which considers three different dimensions of understandability separately, and coarse-grained understandability, which measures the average across the dimensions. The dimensions are:
\begin{description}
    \item[UNC] Understandability not in context  
    measures the comprehensibility of ADTs \textit{syntax}. It assesses users' ability, after ADT training, to identify correct ADT construction, recognise nodes (for attack and defense), refinements (conjunctive and disjunctive), countermeasures, and understand sequential actions and their temporal order in ADTs.
    \item[UIC] Understandability in context  
    measures the comprehensibility of ADTs \textit{semantics}. It assesses users' ability, after training, to answer questions about both existing and instantiated ADTs and to recognise if an ADT accurately models a specific behaviour in a given scenario.
    \item[TRF] Transferability  
    measures the practical use of the notation, evaluating users' ability, after training, to create or modify ADTs. This includes recognising the appropriate elements to add to the tree for modelling specific behaviour and knowing where to place these elements.
\end{description}

\paragraph{Research Questions.} 
We aim to answer the following research questions: 
\begin{description}
    \item[RQ1] \textit{How well users understand ADTs?} This RQ aims to understand the level of effectiveness and efficiency with which users comprehend ADTs.
    \item[RQ2] \textit{What is the degree of acceptance of ADTs by users?} This RQ aims to understand how much users perceive the notation as easy to use and useful and to what extent they intend to use ADTs in the future.
    \item[RQ3] \textit{What is the relationship between ease of use/usefulness of the notation and intention to use it in the future?} Differently from RQ1, which focuses on each perception-based variable independently, this RQ aims at checking whether there is a relationhip among the variables, and in particular, if ease of use and usefulness are related to intention to use. 
    \item[RQ4] \textit{What is the relationship between the overall ADT understandability and the users’ perception of ADTs’ ease of use and usefulness?}
    With this RQ, we check whether users who perform best in understanding the notation also tend to evaluate the ADTs as easier and more useful.
    \item[RQ5] \textit{What is the relationship between the different dimensions of understandability and the users' perception of ADTs' ease of use and usefulness?} Here, we want to check if there is an understandability dimension that is related to the perception of users in terms of ease of use and usefulness. 
\end{description}

\paragraph{Variables for Acceptance and Understandability.}
We measure the three per\-cep\-tion-based variables (PEOU, PU, and ITU) through an instrument adapted from MEM \cite{moody2001dealing}, namely a questionnaire composed of a set of statements for each variable.
We shuffle the statements and add their negated version to avoid systematic response bias (i.e., both the statements ``ADTs are easy to learn" and ``ADTs are not easy to learn" are present)~\cite{abrahao2011evaluating}. 
Users need to evaluate each statement on a Likert scale from 1 (\textit{strongly agree}) to 5 (\textit{strongly disagree}). Table~\ref{tab:perc-stat} shows the list of positive statements for PEOU, PU, and ITU.
Each variable is computed as the mean of its statements points (the points for negative statements are counted as~6 minus the points given as the answer).

\begin{table}[t]
\renewcommand{\arraystretch}{1.2}
\caption{Perception-based statements (positive statements).}
\smallskip
\begin{scriptsize}
\begin{tabular}{p{.94cm}| r  p{10.74cm}} \hline
%\hline
& & Statements \\
\hline
PEOU & 1. & It was easy for me to understand what the ADTs represented. \\
 & 2.  & ADTs are simple and easy to understand. \\
&   3.  &ADTs are easy to learn.\\
 & 4. & Overall, the ADTs were easy to use. \\ \hline

PU & 1. & Overall, I think that ADTs provide an effective means for describing security threats and countermeasures. \\
 & 2. & I believe that ADTs have enough expressiveness to represent security threats and countermeasures.\\
& 3. & Overall, I find ADTs to be useful. \\
& 4. & I believe that ADTs are useful for representing security threats and countermeasures. \\
 & 5. & Using ADTs would improve my performance in describing security threats and countermeasures.\\
 & 6. & I believe that ADTs are organised, clear, concise, and unambiguous.\\
& 7. & I believe the use of ADTs would reduce the time required to represent security threats and countermeasures.\\ \hline

ITU & 1. & If I were to work for a company in the future, I would use ADTs to specify security threats and countermeasures. \\
 & 2. & I intend to use ADTs in the future if given the opportunity.\\
& 3. & I would recommend the use of ADTs to security practitioners.\\
 & 4. & It would be easy for me to become skilled in using ADTs.\\
%\hline
\hline
\end{tabular}
\label{tab:perc-stat}
\end{scriptsize}
\end{table}

Understandability dimensions are measured through specific tasks: 
\begin{enumerate}
    \item %Understandability not in context 
    UNC is measured through a set of true/false questions on domain-agnostic ADT fragments (A, B, C instead of names), to ensure that users' responses are not influenced by knowledge of the domain.
    \item %Understandability in context 
    UIC is evaluated through a set of yes/no questions on instantiated ADTs fragments.
    \item %Transferability 
    TRF is measured through a number of instantiated ADTs fragments to extend with a set of requests.
\end{enumerate}
For each of these dimensions, we compute effectiveness as the number of correct answers over the number of questions and efficiency as effectiveness over time~\cite{abrahao2011evaluating}. Therefore, we have six different variables: \textit{UNC effectiveness}, \textit{UNC efficiency}, \textit{UIC effectiveness}, \textit{UIC efficiency},
\textit{TRF effectiveness}, and  \textit{TRF efficiency}. 
For what concerns total understandability, we compute \textit{understandability effectiveness} as the mean of the effectiveness of the three dimensions and \textit{understandability efficiency} as the mean of the efficiency of the three dimensions.

\paragraph{Hypothesis Testing.}
To answer the research questions, we test a number of NULL hypotheses (cf.\ Table~\ref{table:tests}). Not all the combinations of variables are considered, following the approach by Abraõ~\cite{abrahao2011evaluating}, who relates ITU to PU and PEOU only, and not to performance-based variables.
\begin{table}[t]
\renewcommand{\arraystretch}{1.2}
\caption{Hypotheses for each research question.}
\smallskip
\centering
\scriptsize
\resizebox{\textwidth}{!}{%
\begin{tabular}{|l|l|l|}
\hline
\multirow{2}{*}{RQ1} & \textbf{H}$\bm{1_0}$ & Users are not effective in understanding ADTs \\ \cline{2-3}
 &     \textbf{H}$\bm{2_0}$& Users are not efficient in understanding ADTs \\  \hline
\multirow{3}{*}{RQ2} & \textbf{H}$\bm{3_0}$ & ADTs are perceived as difficult to use \\ \cline{2-3}
 &     \textbf{H}$\bm{4_0}$& ADTs are perceived as not useful \\ \cline{2-3}
 &     \textbf{H}$\bm{5_0}$ & There is no intention to use the ADT in the future \\ \hline
\multirow{3}{*}{RQ3} &    \textbf{H}$\bm{6_0}$ & There is no relationship between perceived ease of use and perceived usefulness  \\ \cline{2-3}
 &    \textbf{H}$\bm{7_0}$ & There is no relationship between perceived usefulness and intention to use \\ \cline{2-3}
 &    \textbf{H}$\bm{8_0}$ & There is no relationship between perceived ease of use and intention to use   \\ \hline   
\multirow{4}{*}{RQ4} &    \textbf{H}$\bm{9_0}$ & There is no relationship between understandability effectiveness and perceived ease of use      \\ \cline{2-3}
 &     \textbf{H}$\bm{10_0}$ & There is no relationship between understandability effectiveness and perceived usefulness \\ \cline{2-3}
 &     \textbf{H}$\bm{11_0}$ & There is no relationship between understandability efficiency and perceived ease of use   \\ \cline{2-3}
 &     \textbf{H}$\bm{12_0}$ & There is no relationship between understandability efficiency and perceived usefulness  \\ \hline  
\multirow{12}{*}{RQ5} &     \textbf{H}$\bm{13_0}$ & There is no relationship between understandability not in context effectiveness and perceived ease of use  \\ \cline{2-3}
 &     \textbf{H}$\bm{14_0}$ & There is no relationship between understandability not in context effectiveness and perceived usefulness \\ \cline{2-3}
 &     \textbf{H}$\bm{15_0}$ & There is no relationship between understandability not in context efficiency and perceived ease of use \\ \cline{2-3}
 &     \textbf{H}$\bm{16_0}$ & There is no relationship between understandability not in context efficiency and perceived usefulness \\ \cline{2-3}
 &     \textbf{H}$\bm{17_0}$ & There is no relationship between understandability in context effectiveness and perceived ease of use  \\ \cline{2-3}
 &     \textbf{H}$\bm{18_0}$ & There is no relationship between understandability in context effectiveness and perceived usefulness  \\ \cline{2-3}  
 &     \textbf{H}$\bm{19_0}$ & There is no relationship between understandability in context efficiency and perceived ease of use \\ \cline{2-3}
 &     \textbf{H}$\bm{20_0}$ & There is no relationship between understandability in context efficiency and perceived usefulness \\ \cline{2-3}  
 &     \textbf{H}$\bm{21_0}$ & There is no relationship between transferability effectiveness and perceived ease of use    \\ \cline{2-3}
 &     \textbf{H}$\bm{22_0}$ & There is no relationship between transferability effectiveness and perceived usefulness  \\ \cline{2-3}  
 &     \textbf{H}$\bm{23_0}$ & There is no relationship between transferability efficiency and perceived ease of use  \\ \cline{2-3}
 &     \textbf{H}$\bm{24_0}$ & There is no relationship between transferability efficiency and perceived usefulness  \\ \hline
\end{tabular}
}
\label{table:tests}
\end{table}

\subsection{Study Phases}\label{sec:phases}

The study is conducted online (material  in~\cite{zenodo10067064}) and structured in 6 phases. 

\textit{Phase 1 -- Recruitment.}
Participants are contacted through a recruitment e-mail with all the information needed to perform the study. Specifically, links to a video training, a spreadsheet file where to get their identifier and the link to the test, the pre- and post-test questionnaires, the consent form, and study instructions.

\textit{Phase 2 -- Binding.}
To ensure anonymity, participants are provided with a unique alphanumeric identifier via a spreadsheet file with a link to their test document (there is a different document for each participant). They are instructed to keep the identifier for the entire test, preserve the link to the test document to be used in a subsequent phase, and use incognito mode to protect their identity.

\textit{Phase 3 -- Training.}
Before starting the test, we ask participants to watch a video that presents the ADT notation. The video is 
available online (\url{https://youtu.be/KLIH-yultgI})
and it
contains all the information needed to complete the test. 
Participants are asked to  
use this support only once before they begin the test. 

\textit{Phase 4 -- Pre-test questionnaire.}
We ask participants to fill out an online questionnaire whose link has been sent by e-mail during the recruiting phase. 
The questionnaire collects information about gender, age, education, employment, work area, level of knowledge of ADTs, and education on ADTs. 
Participants have to mark the questionnaire with the identifier received during the binding phase (Phase~2).

\textit{Phase 5 -- Test.}
We ask participants to fill out the test in all its phases. The test is accessible through the link received during the binding phase (Phase~2); such a  link leads to an editable online document (a different document for each participant). The spreadsheet accessed in Phase~2 enables us to bind each document to the ID of the corresponding user.
The test is composed of 4~steps: 
\begin{enumerate}[i]
    \item \textbf{Retention.} 
Retention measures the comprehension of the training material and the ability to retain knowledge from it. We use this step to keep in the participants' memory the concepts presented in the training video that they will need during the test. The outcome of this step is not utilised in the calculation of understandability.
In this step, a list of figures (i.e., all figures in the legend of Fig.~\ref{fig:ADT}) is presented and, for each figure, a table with two definition options. Participants are asked to mark the right definition for each figure. 
\item \textbf{Understandability not in context.}
With this step, we want to get how understandable is the syntax of the notation for the participants.
In this step, 6~items are presented, and for each of them, we show one or more attack-defense tree fragments and 4 statements. Participants have to check for each of the statements whether it is true or false.
Participants are asked to write down the starting (when starting step~ii) and finishing time (when completing all the steps).
\item \textbf{Transfer.}
Transfer measures how much is transferable the knowledge acquired through the training material. 
In this step, three attack-defense tree fragments are presented and, for each of them, a list of three requests. Participants are asked to modify the tree fragments according to the requests using an editable 
diagram embedded in the 
document (the instructions to modify the diagram are written inside the diagram itself).
The three ADT fragments used represent common and familiar types of attacks, namely an attack on a bank account, an attack to open a safe lock, and an attack to burgle a house. For each fragment, three requests were made, each with increasing levels of difficulty: (i)~participants are asked to add a node to the tree and specify the type of node and its position; (ii)~participants are asked to add all the nodes necessary to model a given situation; (iii)~participants are asked to modify the tree according to given syntactic and/or semantic constraints.  
For each of the three items, participants are asked to write down starting and finishing times in the appropriate lines.
\item \textbf{Understandability in context.}
With this step, we want to perceive to what extent users, after a training phase on ADTs, are able to answer questions about given ADTs. 
In this step, three attack-defense tree fragments are presented, and, for each of them, a list of three yes/no questions. Participants are asked to answer the questions by typing in the document ``yes" or ``no". 
The three ADT fragments used are extended versions of the fragments used in the Transfer step (cf. step iii).
For each of the three items, participants are asked to write down starting and finishing times in the appropriate lines.
\end{enumerate}

Users are not bound by a specific time frame for the test phase, but allocating 40 minutes is deemed sufficient for completing phases ii, iii, and iv (according to the authors ter Beek and Lluch Lafuente, who are ADT experts \cite{terbeek2021quantitative}). This duration considers the time required for reading and analysing questions, processing ADT fragments, providing accurate answers, and adapting to the platform used.

%Users are not required to complete the test phase within a specific time frame. However, we consider it adequate to allocate 40 minutes to complete phases ii, iii, and iv (indicating a sufficient performance time according to the authors ter Beek and Lluch Lafuente, who are ADT experts \cite{terbeek2021quantitative}). 
%This time frame takes into account the time needed to read and analyse the questions, process the ADT fragments, and answer the questions accurately, as well as the time to get used to the platform used.
\medskip

\textit{Phase 6 -- Post-test questionnaire.}
We ask participants to fill out an online questionnaire whose link has been sent by e-mail during the recruiting phase.  
We use this phase to measure the perception-based variables (namely, PEOU, PU, and ITU) through a set of statements users need to rate from 1 to 5. The questionnaire contains 8 statements concerning PEOU, 14 statements on PU, and 8 statements concerning ITU (see Table \ref{tab:perc-stat}).
Participants have to mark the questionnaire with the identifier received during the binding phase (Phase~2).

\section{Study Execution}
\label{sec:execution}

The experimental study protocol containing the definition of the study phases, its rationale, as well as the data analysis process has been submitted to the ethical committee of the Italian National Research Council (CNR), which authorised the administration of the test.
To take part in the study, participants are asked to sign an informed consent for the processing of personal data.

\begin{figure}
\centering
\begin{minipage}{.5\textwidth}
  \centering
  \includegraphics[width=.63\linewidth]{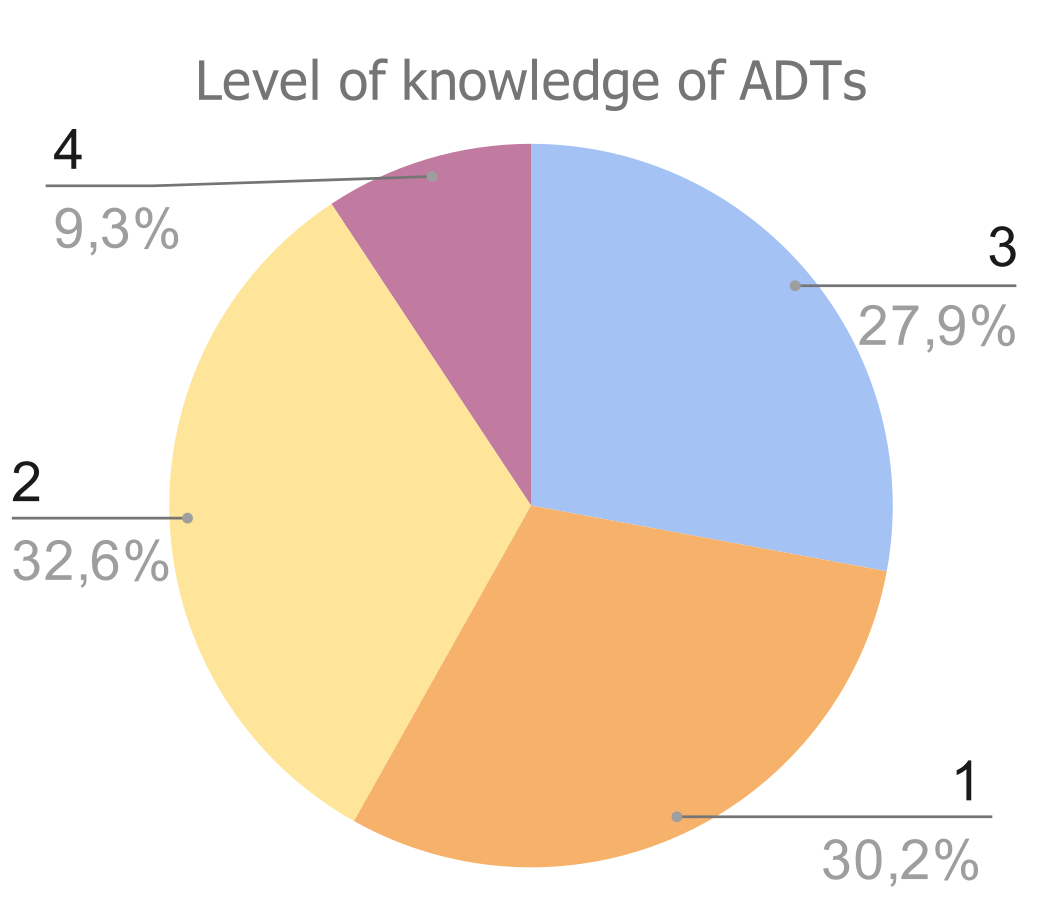}
  \captionof{figure}{Level of knowledge of ADTs.}
  \label{fig:pieChart1}
\end{minipage}%
\begin{minipage}{.5\textwidth}
  \centering
  \includegraphics[width=1\linewidth]{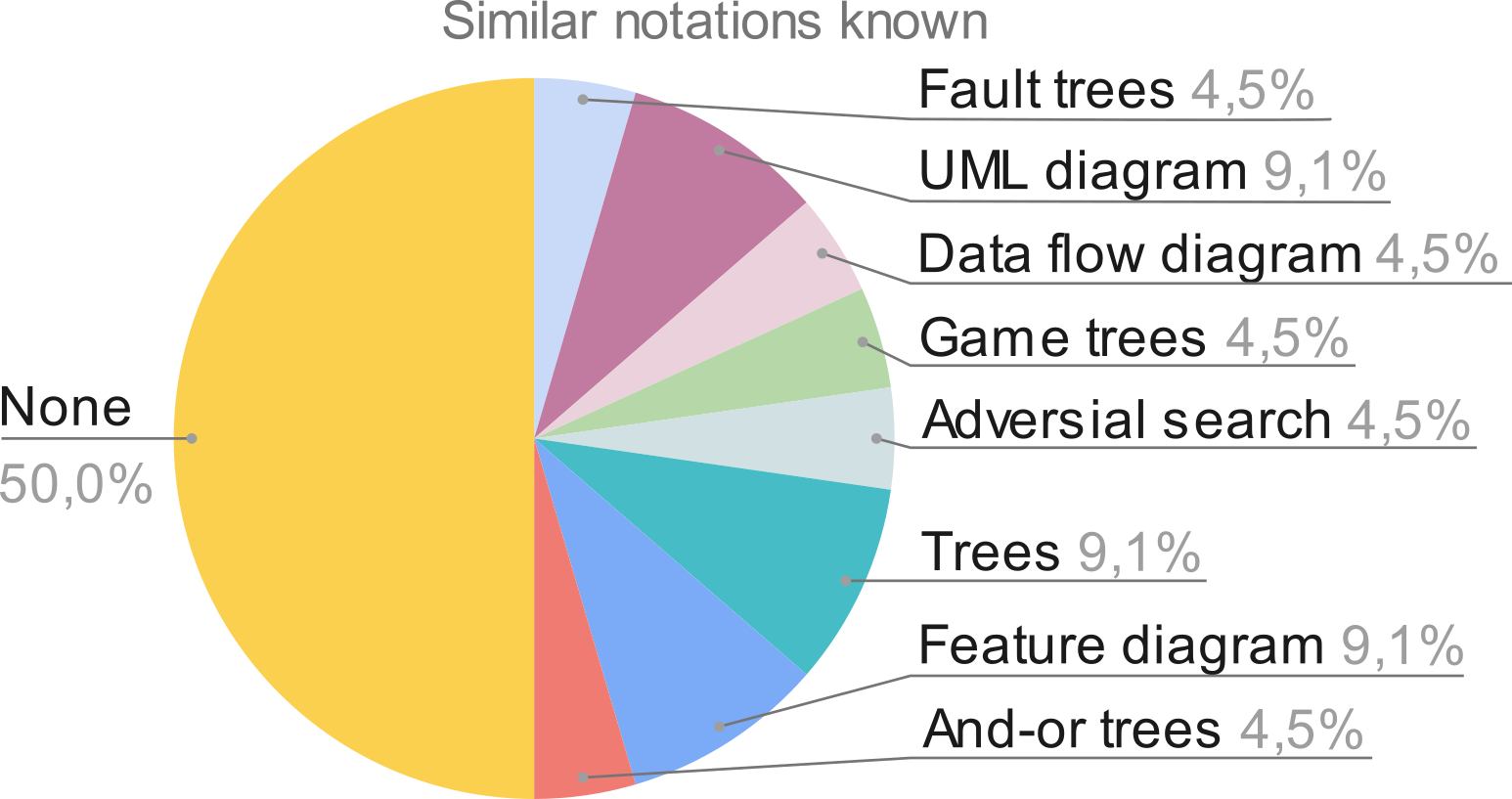}
  \captionof{figure}{Similar notations known.}
  \label{fig:pieChart2}
\end{minipage}
\end{figure}

\paragraph{Participants.}
In total, 25~participants took part in the study: computer science students, Ph.D. students, and professors; researchers in the field of software engineering, formal methods, and security; participants belong to Kennesaw State University, CNR, University of Pisa, and the Technical University of Denmark. Participants in the study were selected opportunistically based on their availability.
They were of both genders (56\%~men, 40\%~women, 4\%~prefers not to answer), aged between~21 and~56 years old.
We asked them to self-evaluate their knowledge of ADTs before the test on a 5-point scale from~1 (\textit{no knowledge}) to~5 (\textit{advanced}) and whether they knew similar notations. The results are reported in Figures~\ref{fig:pieChart1} and~\ref{fig:pieChart2}, respectively.
A total of 80\% of the participants did not receive any education on ADTs before the test; the remaining participants attended a university course, a seminar, or self-educated.

\paragraph{Results.}
Table~\ref{tab:descrStats} shows descriptive statistics for all the variables gathered with the test, i.e., the perception-based variables (PEOU, PU, and ITU) and the performance-based variables: (1)~understandability not in context effectiveness and (2)~efficiency; (3)~understandability in context effectiveness and (4)~efficiency; (5)~transferability effectiveness and (6)~efficiency; and (7)~understandability effectiveness and (8)~understandability efficiency. 

\begin{table}[t]
    \renewcommand{\arraystretch}{1.2}
    \caption{Descriptive statistics.}
    \label{tab:descrStats}
    \smallskip
    \centering
    \scriptsize
    \vspace*{-0.05cm}
    \begin{tabular}{|c | c| c| c |c | c|}
    \hline 
      {\bf{\em Variables}} & {\bf{\em Median}}  & {\bf{\em Mean}} & {\bf{\em Std.\,dev.}} & {\bf{\em Min.}} & {\bf{\em Max.}} \\ 
      \hline \hline
    PEOU & 4.25  & 4.18 & 0.563 & 2.875 & 5\\ \hline
    PU & 4  & 3.92 & 0.37 & 2.929 & 4.571 \\  \hline 
    ITU &  3.875 & 3.88 & 0.403 & 3 & 5 \\ \hline \hline
    UNC effectiveness & 0.750 & 0.783 & 0.083 & 0.625 & 0.958 \\ \hline
    UNC efficiency & 0.094 & 0.103 & 0.046 & 0.024 & 0.188 \\ \hline
    UIC effectiveness & 0.889 & 0.907 & 0.175 & 0.111 & 1 \\ \hline
    UIC efficiency & 0.250 & 0.264 & 0.135 & 0.009 & 0.500 \\ \hline
    TRF effectiveness & 0.667 & 0.613 & 0.267 & 0 & 1 \\ \hline
    TRF efficiency & 0.023 & 0.026 & 0.015 & 0 & 0.049 \\ \hline
    understandability effectiveness & 0.792 & 0.768 & 0.134 & 0.287 & 0.986 \\ \hline
    understandability efficiency & 0.118 & 0.131 & 0.059 & 0.011 & 0.241 \\ \hline
   \end{tabular}
\end{table}

The perception-based variables 
are all above the average value of the Likert scale (i.e., 3), which thus suggests a general degree of acceptance of the notation. 
The results indicate that users perceive ADTs as easy to use (mean score of~4.18) and as useful (mean value~3.92). Results also suggest that users intend to use the notation in the future (ITU has an average score of~3.88).

The results indicate a generally good level of understandability of ADT notation with an average total understandability effectiveness above~0.76, meaning that $\sim77\%$~of the questions of the test are correctly answered.

Regarding the different dimensions composing understandability, the results show that understandability in context is the measure that provides the highest contribution (average effectiveness of~0.907), followed by understandability not in context (effectiveness~=~0.783) and transferability (effectiveness~=~0.613). This suggests that while participants understand the syntax and semantics of ADT fragments, they have more difficulty applying them in practice.
For what concerns efficiency, we observe a similar trend, thereby confirming that ADTs ``in action" are perceived as more difficult.

Table~\ref{tab:hyp} summarises the relation between variables expressed in the hypotheses addressing each RQs presented in Section \ref{sec:stDes}. For each hypothesis, the column ``Reject" reports a ``T" if the hypothesis has been rejected and an ``F" otherwise. 
Below we discuss in detail only the rejected NULL hypotheses because no conclusions can be made for the others.

\begin{table}[t]
    \renewcommand{\arraystretch}{1.2}
    \caption{Statistics summary.\\Blue rows indicate NULL hypotheses that have been rejected ({\upshape p}-value $<$ 0.05)\\
    The term ``effv" indicates effectiveness and the term ``effc" indicates efficiency.}
    \label{tab:hyp}
    \smallskip
    \centering
    \scriptsize
   % \vspace*{-0.15cm}
    \resizebox{\textwidth}{!}{%
    \begin{tabular}{|c | c | c| c|c |c | }
    \hline 
      {\bf{\em RQs}} & {\bf{\em Hyp.}} & {\bf{\em Variables}}  & {\bf{\em Reject}}&  {\bf{\em p-value}} & {\bf{\em Effect-size}} \\ \hline 
\multirow{2}{*}{RQ1} & \cellcolor{blue!25}H$1_0$ &  \cellcolor{blue!25}{Effectiveness} & \cellcolor{blue!25}T  & \cellcolor{blue!25}$0.000139$ & \cellcolor{blue!25} $1.255714$ \\      
      &  \cellcolor{blue!25}H$2_0$ &  \cellcolor{blue!25}{Efficiency} & \cellcolor{blue!25}T  & \cellcolor{blue!25}$7.381E-06$ & \cellcolor{blue!25}$1.983621$
 \\   \hline

      \multirow{3}{*}{RQ2} & \cellcolor{blue!25}H$3_0$ &  \cellcolor{blue!25}{PEOU} & \cellcolor{blue!25}T  & \cellcolor{blue!25}$1.08E-05$ & \cellcolor{blue!25}$2.097433$\\      
      &  \cellcolor{blue!25}H$4_0$ &  \cellcolor{blue!25}{PU} & \cellcolor{blue!25}T  & \cellcolor{blue!25}$7.11E-06$& \cellcolor{blue!25}$2.485847$\\ 
      &  \cellcolor{blue!25}H$5_0$ &  \cellcolor{blue!25}{ITU} & \cellcolor{blue!25}T  & \cellcolor{blue!25}$9.28E-06$& \cellcolor{blue!25}$2.185815$\\ \hline\hline
{\bf{\em RQs}} & {\bf{\em Hyp.}} & {\bf{\em Relation between variables}}  & {\bf{\em Reject}}& {\bf{\em Eq.}}& {\bf{\em p-value}} \\ \hline
     \multirow{3}{*}{RQ3} & H$6_0$ &  PEOU $\rightarrow$ PU & F & PU = 3.6 + 0.073 * PEOU & 0.5962\\
      &  \cellcolor{blue!25}H$7_0$ &  \cellcolor{blue!25}{PU $\rightarrow$ ITU} & \cellcolor{blue!25}T & \cellcolor{blue!25}ITU = 0.5 + 0.86 * PU & \cellcolor{blue!25}2.44E-06\\  
      & H$8_0$ &  PEOU $\rightarrow$ ITU & F &ITU = 2.9 + 0.24 * PEOU
 & 0.108\\ \hline
      
      \multirow{4}{*}{RQ4} & \cellcolor{blue!25}H$9_0$ &  \cellcolor{blue!25}{und.\,effv $\rightarrow$ PEOU} & \cellcolor{blue!25}T & \cellcolor{blue!25}{PEOU = 2.8 + 1.8 * und.\,effv} &  \cellcolor{blue!25}0.03677\\     
      &  H$10_0$ &  und.\,effv $\rightarrow$ PU & F & PU = 3.2 + 0.97 * und.\,effv & 0.08483\\  
      & H$11_0$ &  und.\,effc $\rightarrow$ PEOU & F & PEOU = 3.8 + 2.7 * und.\,effc & 0.1752 \\      
      &  H$12_0$ &  und.\,effc $\rightarrow$ PU & F & PU = 4 - 0.5 * und.\,effc & 0.8492\\  \hline

       \multirow{12}{*}{RQ5} 
       %UNDERSTANDABILITY NOT IN CONTEXT
      &  H$13_0$ &  UNC effv $\rightarrow$ PEOU & F & PEOU = 4.5 - 0.74* UNC effv & 0.7578 \\ 
      &  H$14_0$ &  UNC effv $\rightarrow$ PU & F & PU = 4.5 - 0.44 * UNC effv & 0.4241\\  
      & H$15_0$ &  UNC effc $\rightarrow$ PEOU & F & PEOU = 3.9 + 2.9 * UNC effc & 0.2606 \\   
      & H$16_0$ &  UNC effc $\rightarrow$ PU & F & PU = 3.9 - 0.22 * UNC effc
 & 0.8952\\
     %UNDERSTANDABILITY IN CONTEXT
      & \cellcolor{blue!25}H$17_0$ &  \cellcolor{blue!25}{UIC effv $\rightarrow$ PEOU} & \cellcolor{blue!25}T & \cellcolor{blue!25}PEOU = 2.8 + 1.5 * UIC effv & \cellcolor{blue!25}0.02051 \\  
      &  H$18_0$ &  UIC effv $\rightarrow$ PU & F & PU = 3.5 + 0.43 * UIC effv
 & 0.3332\\
      & H$19_0$ &  UIC effc $\rightarrow$ PEOU & F & PEOU = 3. 9 + 1.1 * UIC effc & 0.2168\\   
      & H$20_0$ &  UIC effc $\rightarrow$ PU & F & PU = 4 - 0.16 * UIC effc
 & 0.7812\\  
    %TRANSFERABILITY
      &  H$21_0$ &  {TRF effv $\rightarrow$ PEOU} & F & PEOU = 3.7 + 0.73 * TRF effv
        & 0.08802\\ 
        &  \cellcolor{blue!25}H$22_0$ &  \cellcolor{blue!25}{TRF effv $\rightarrow$ PU} & \cellcolor{blue!25}T & \cellcolor{blue!25}PU = 3.5 + 0.62 * TRF effv
 & \cellcolor{blue!25}0.02494\\  
      & H$23_0$ &  TRF effc $\rightarrow$ PEOU & F &PEOU = 3.9 + 10 * TRF effc & 0.2105 \\
      & H$24_0$ &  TRF effc $\rightarrow$ PU & F & PU = 3.8 + 3.9 * TRF effc
& 0.4685\\
      \hline
   \end{tabular}
   }
\end{table}

\textbf{RQ1.} To answer RQ1, we applied a Wilcoxon signed rank test to check whether effectiveness and efficiency are significantly above the target values of~0.6 (indicating a sufficient performance according to ter Beek and Lluch Lafuente, two ADT experts \cite{terbeek2021quantitative}) and of~0.015 (i.e., 60\% of~1 (maximum effectiveness)/40 min (expected completion time)), respectively. We apply a  non-parametric test (i.e. Wilcoxon signed rank) because the normality check, performed with the Kolmogorov-Smirnov test,  fails for all the variables with p-value well below the 0.05 significance level.
The test results show that both variables are significantly higher than the target values for $\alpha~=~0.05$, with p-values of~0.000139 and~7.381e-06, respectively, with large effect-size (cf.\ Table~\ref{tab:hyp}). Therefore rejecting H$1_0$ and H$2_0$, and attesting a \textbf{sufficient overall understandability of the ADT notation}. Fine-grained effectiveness and efficiency measures are all significantly greater than the respective reference values for both effectiveness and efficiency, with the exception of transferability; we refer to~\cite{zenodo10067064} for detailed information.

\textbf{RQ2.} To answer RQ2, we applied a Wilcoxon signed rank test to check whether PEOU, PU, and ITU are significantly above the average value of the Likert scale (i.e., 3).
The test results show that all the variables attesting the acceptance are significantly higher than~3 for $\alpha~=~0.05$, with p-values of~1.077e-05, 7.109e-06, and~9.282e-06, respectively, with large effect-size (cf.\ Table \ref{tab:hyp}). Therefore rejecting H$3_0$, H$4_0$, and H$5_0$  and confirming the overall degree of acceptance of the ADT notation as high.
\begin{figure}[ht!]
\centering
%\subfloat[Boxplot of users' acceptance variables]
%{% Input figure 1
\includegraphics[height=4.5cm]{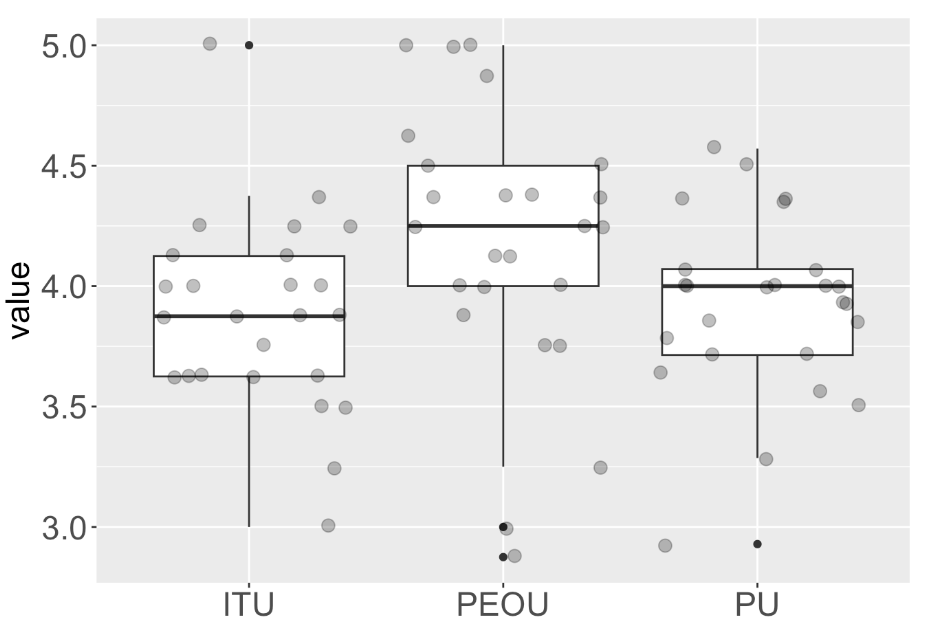}
%\label{fig:RQ1}
%}
\caption{\label{fig:RQ1}Result for RQ2: boxplot of users' acceptance variables}
\end{figure}
\noindent
As the boxplot in Figure~\ref{fig:RQ1} shows, while ITU and PU have comparable values, PEOU receives the highest score. This suggests that \textbf{ease of use is the main characterising quality of ADTs}.

\begin{figure}
\centering
\subfloat[Relationship between PEOU and \\PU and ITU.]
{% Input figure 2
\includegraphics[height=4.25cm]{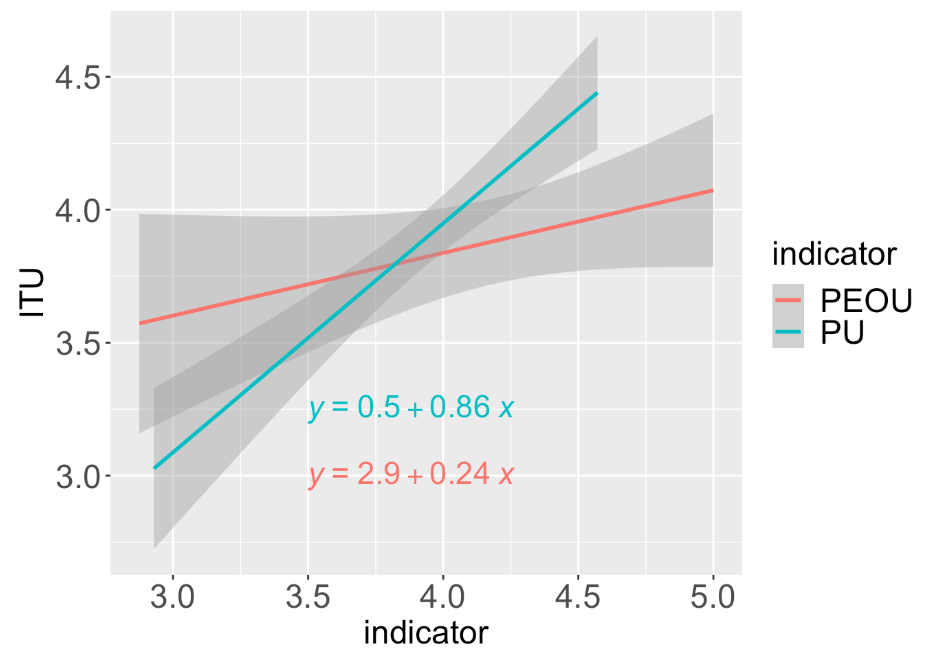}
\label{fig:RQ2}
}
\centering
\subfloat[Relationship between understandability\\ effectiveness and PEOU and PU.]
{% Input figure 1
\includegraphics[height=4.25cm]{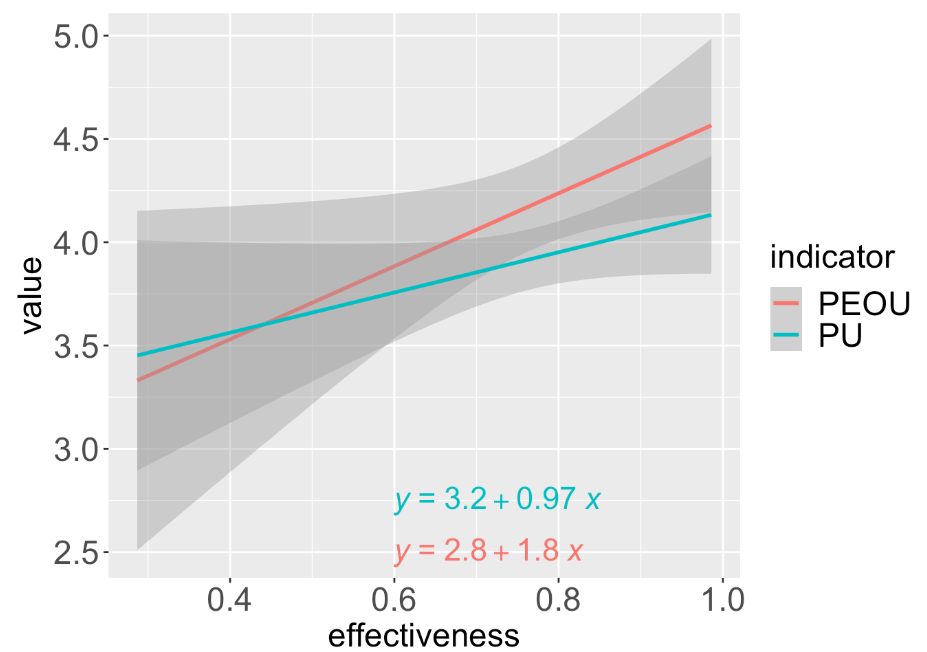}
\label{fig:RQ3}
}
\caption{Results for RQ3 and RQ4.}
\end{figure}

\textbf{RQ3.} To answer RQ3, we fit a regression linear model between PEOU and PU, and between both PEOU and PU and ITU. As shown in Figure~\ref{fig:RQ2}, the test results attest that there is a significant positive relationship between PU and ITU (p-value = 2.44e-06). We can thus reject H$7_0$ and suggest that \textbf{users intend to use the notation in the future more for its usefulness than for its easiness}.

\textbf{RQ4.} 
To check if there is a relationship between the understandability of the notation and the users' perceptions about its easiness and usefulness, we test  H$9_0$--H$12_0$ by fitting a linear model between PEOU and PU and understandability effectiveness, and between PEOU and PU and understandability efficiency.
Our results show 
a significant positive relationship between effectiveness and perceived ease of use (cf.\ Fig.~\ref{fig:RQ3}), suggesting that \textbf{users who perform best in the test tend to evaluate better the notation in terms of easiness}.

\begin{figure}
\centering
\subfloat[Relationship\,between\,understandability\\ in context effectiveness and PEOU and PU.]
{% Input figure 2
\includegraphics[height=4.15cm]{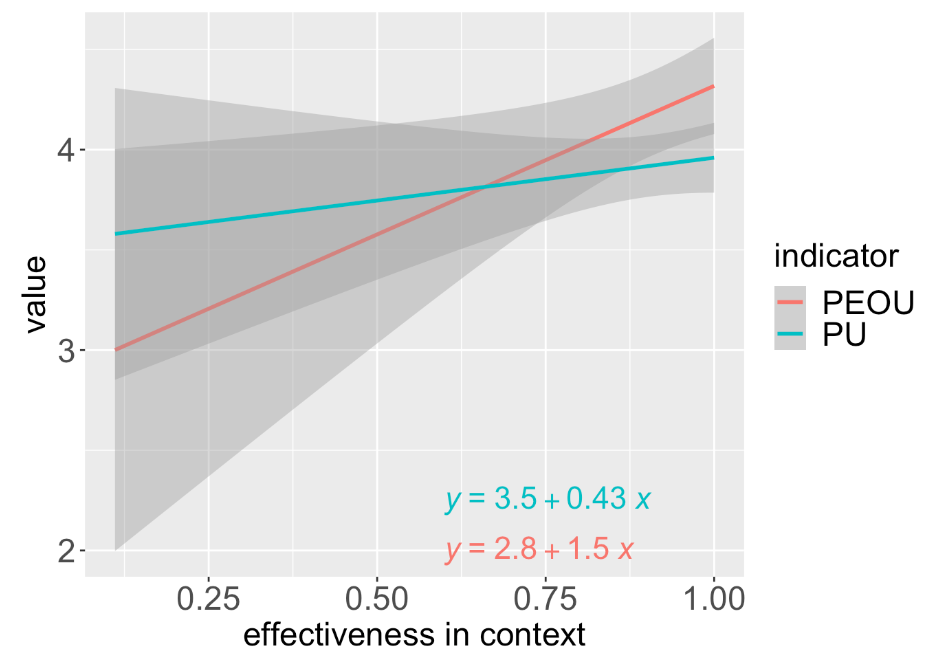}
\label{fig:RQ4_1}
}
\,\
\centering
\subfloat[Relationship between transferability\\ effectiveness and PEOU and PU.]
{% Input figure 2
\includegraphics[height=4.15cm]{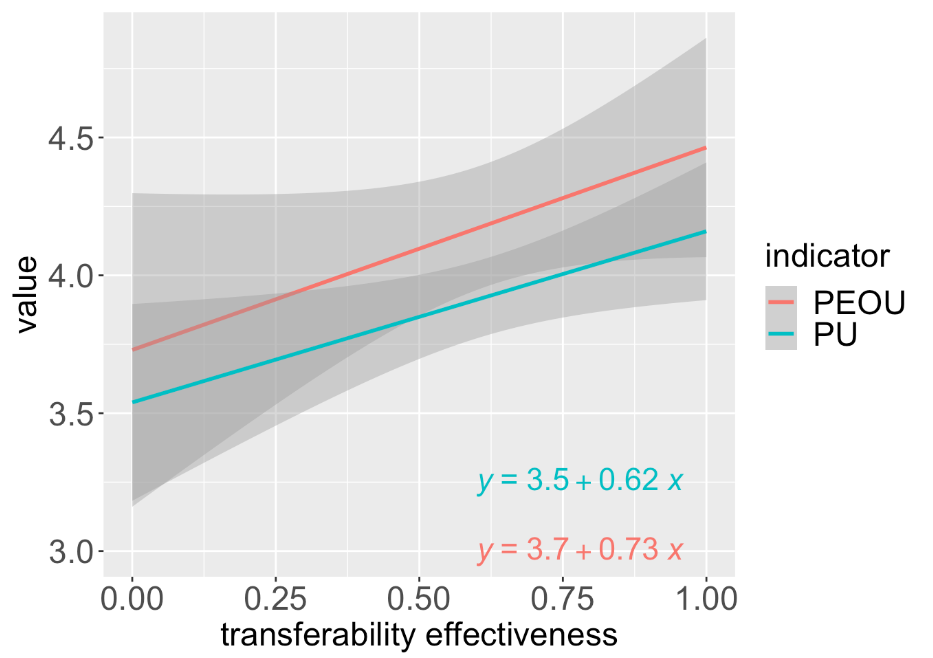}
\label{fig:RQ4_2}
}
\caption{Results for RQ5.}
\end{figure}

\textbf{RQ5.} Finally, to understand whether one of the understandability dimensions affects most the perceived easiness and usefulness,
we fit a regression linear model between the perception-based variables (PEOU and PU), and the effectiveness and efficiency of the three understandability dimensions (understandability not in context, understandability in context, and transferability). 
Our results show that the effectiveness of understandability in context and of transferability both have a significant positive relationship
with PEOU and PU, respectively (cf.\ Figs.~\ref{fig:RQ4_1} and~\ref{fig:RQ4_2}).
We can thus reject H$17_0$ and H$22_0$, and confirm that 
users who observed instantiated trees and understand their meaning tend to evaluate the notation as easier, while
users who apply the method better by extending the tree correctly tend to evaluate it as more useful.
This suggests that \textbf{users~who successfully use the notation in practice, tend to appreciate it more}. 

\subsection{Threats to Validity}

\textit{Construct Validity.} 
Users' acceptance was assessed through existing models~\cite{moody2001dealing} and adapted to the ADT notation according to \cite{abrahao2011evaluating}.
The usage of effectiveness and efficiency for understandability performance is widely used in the literature (cf., e.g., \cite{abrahao2011evaluating,moody2001dealing,broccia23neverlang}). 
For what concerns the understandability dimensions, retention and transferability are adapted from \cite{mayer1989models,abrahao2011evaluating}, even if here we use retention as a means to retain the information gathered from the training phase rather than a dimension to be measured.
Understandability not in context and in context are measures adapted from \cite{abrahao2011evaluating} to address the evaluation of syntax and semantics.
The tasks used for each dimension have been revised by two ADT experts and considered appropriate to evaluate the understandability of the notation.

\textit{Internal Validity.}
To prevent systematic response bias in user acceptance questionnaires, we mixed positive and negative statements.
Moreover, to minimise participant response bias and limit the possible tendency of users to provide positive answers to please the researchers, the experiment was conducted completely online; thus, none of the users met the experimenter. This approach not only preserved participant anonymity but also created a more naturalistic setting, minimising biases introduced by participants' awareness of being observed and diminishing the Hawthorne Effect.
The support used during the test (e.g., the editable online document and diagram) may have influenced users' performance. To study this hypothesis, further investigation with users must be carried out to grasp their difficulties with the support.

\textit{External Validity.}
%The study's problem-solving tasks mirror typical actions performed by ADT users, such as comprehending tree meaning and syntax, reading existing trees, and creating or extending ADTs. We deem these tasks sufficiently representative to evaluate understandability.
 %Other tasks (e.g., evaluating the attack/defense cost with statistical model checking~\cite{terbeek2021quantitative}) were excluded, potentially yielding different outcomes. 
The selected participants encompass diverse genders and experience levels, enhancing generalisability. Participants were opportunistically chosen from the academic field, varying in seniority. However, their representation may not fully encompass all ADT user classes, influencing study results. Further research involving users from different fields is needed to confirm the applicability of conclusions across all user classes. It should also be noted that this study is a controlled experiment, which aims to maximise internal validity and does not evaluate ADT users in a realistic setting, where contextual factors play a relevant role. Therefore, case studies are needed to confirm that our conclusions apply in a real-life security analysis environment. 

%The problem-solving tasks proposed in the study reproduce some typical tasks that ADT users generally perform: understanding the meaning and the syntax of trees, reading an existing tree, and building an ADT from scratch or extending one.
%All other possible tasks (e.g., evaluating the attack/defense cost with statistical model checking~\cite{terbeek2021quantitative}) have not been considered, and different outcomes might be observed with different tasks. Concerning the scope of validity, the selected subjects cover different genders and degrees of experience, ensuring a good generalisability level.
%The participants in the study were chosen opportunistically based on their availability. While of different seniority levels, they all belong to the academic area and are not fully representative of the various ADT classes of users. This lack of representation may have influenced the results of the study. To test this hypothesis, further research involving users from different fields must be conducted to determine whether the conclusions apply to all classes of users.

\section{Conclusion and Future Work}
\label{sec:conclusion}

In this paper, we presented the first empirical study to assess the quality of ADTs in terms of users' acceptance and understandability. Our evaluation measures how well the notation can be used in practice.
In particular, our study focused on assessing users' perceptions variables that attest the notation appreciation in terms of ease of use, usefulness, and intention to use, and of performance variables that attest the degree of understandability of the notation in terms of effectiveness and efficiency. Understandability has also been studied according to three different fine-grained dimensions, and the relation between all these variables has been evaluated through multiple statistical tests.

Our results suggest that the ADT notation is sufficiently understood and greatly appreciated by users, specifically, the main aspect characterising its quality is its ease of use. 
Overall, the notation has a good level of understandability with a total average effectiveness above~0.76. Among its dimensions, we note better performance in more practical tasks (i.e., those related to observing and extending instantiated trees).
Concerning relationships among the variables, we note that general understandability and understandability in context have a relationship with the perceived ease of use and that the ability to apply ADT in practice has a relationship with the perceived usefulness.

In future research, we plan to address user challenges in the test by conducting interviews to assess the impact of the platform on performance. To enhance result accuracy, we will broaden our subject pool, including users from diverse classes, such as those in the security field. We also intend to compare user performance and perceptions across ADTs and other security requirements modelling techniques, preferably textual methods. Additionally, our analysis will encompass various commercial and academic ADT tools.

{\bigskip
\small
\noindent{\bf Acknowledgements.}
Research supported by the Italian MUR--PRIN 2020TL3X8X project T-LADIES (Typeful Language Adaptation for Dynamic, Interacting and Evolving Systems); by Innovation Fund Denmark and the Digital Research Centre Denmark, through the bridge project ``SIOT – Secure Internet of Things – Risk analysis in design and operation"; by Industriens Fond through the project ``Sb3D: Security-by-Design in Digital Denmark"; and by the EU Project CODECS GA 101060179. The authors would like to thank all the participants of the study. 
}

%\clearpage
\bibliographystyle{splncs04}
\bibliography{biblio}

\end{document}